\documentstyle[aps,prb,amsmath]{revtex}
\title{Spatially nonuniform energy distribution of electrons
 in a submicron semiconductor film}
\author{Yu.~G.~Gurevich, G.~N.~Logvinov, and O.~Yu.~Titov}
\address{Institute of Radiophysics and Electronics, Academy 
of Sciences of the Ukraine, 310085, Kharkov, Ukraine}
\date{Submitted October 26, 1992; accepted for publication February 1, 
1993} 
\begin{document}
\preprint{Fiz. Tekh. Poluprovodn. 27, 1040-1046 (June 1993)}
\maketitle

\makeatletter
\global\@specialpagefalse
\def\@oddhead{%
Semiconductors {\bfseries 27} (6), June 1993, p.~566--569
\hfill
\copyright{} 1993 American Institute of Physics}
\let\@evenhead\@oddhead

\begin{abstract} 
The symmetric part of the distribution of the electrons in a 
semiconductor submicron film, placed between a heater and refrigeration
unit, is derived and analyzed. It is shown that, in general, it is of
non-Fermi (non-Maxwellian) nature. A new mechanism is proposed to
account for the non-Maxwellian form of the symmetric part of the
distribution function. This mechanism is based on the different energy
dependences of the momentum relaxation time in the bulk of the
semiconductor and its skin layer.
\end{abstract}

In recent years we have witnessed an increase in the
number of problems in which transport phenomena occur
under conditions such that the flux of charge carriers moves
along an inhomogeneity of the symmetric part of the energy
distribution function, which is caused by a departure of carriers from
equilibrium (due to an external stimulus).

When the symmetric part of the nonequilibrium distribution function of
carriers is of the Fermi (Maxwellian) type, the
problem can be solved relatively simply (see, for example,
Refs.~\onlinecite{SL69,GS72,VG78,BZK82,GY87,GL92PR}) by the use of 
effective boundary conditions.\cite{BBG84}

However, there are frequency cases in which for a variety
of reasons the symmetric part of the equilibrium distribution
function is not of the Fermi type.\cite{IMSK87,GY89,GL92PSS} In this 
case, if the external stimulus is an externally imposed potential
difference, the physical phenomena are generally insensitive to the
actual form of the symmetric part of the distribution 
functional\cite{BG75} (the few exceptions\cite{L65,GLY92} confirm 
the general rule). The situation is different if the electron flux is
caused by the presence of a refrigerator or a heater. We recall that in
the temperature approximation (in which the symmetric part of the
distribution function is Maxwellian) the role of the external force is
played by a temperature gradient. It is necessary in this case to single
out the contribution due to a gradient of the chemical potential $\mu$
(Refs.~\onlinecite{A81} and~\onlinecite{GY91}), i.e., by dividing the 
thermal diffusion flux into a flux which is due to the built-in
thermoelectric field ($\approx\nabla\mu$) and which is compensated by 
the drift flux ($\approx\nabla\varphi$, where $\varphi$ is the 
electrical potential) and a thermal diffusion flux which is included in
the external circuit and which forms the thermoelectric current.

However, if the symmetric part of the distribution function is not of
the Fermi (Maxwellian) type, the absence of the concepts of temperature
gradient and chemical potential gradient (the concepts of temperature
and chemical potential in general) creates the problem of separation of
the thermal diffusion flux. Accordingly, in the absence of temperature
and of chemical potential it is necessary to reformulate the problem of
calculation of the thermoelectric current and the thermoelectric emf.

It is shown in Ref.~\onlinecite{GY91} that a correct calculation of an 
emf of any nature requires consideration of closed circuits, i.e., a
study of the physical phenomena which give rise to an emf in the
presence of a transport current. 

This problem will be dealt with in a separate communication. Our aim
here is to prove the hypothesis that a semiconductor film whose
thickness is less than the characteristic energy relaxation length
(which corresponds to submicron thicknesses) which is placed between a
heater and a refrigerator, and which is short-circuited to an external
resistance generally forms a distribution function of carriers whose
symmetric part is not of the Fermi (Maxwellian) type. We shall show
below that there is a fundamentally new mechanism for the formation of
the symmetric part of the distribution function which is related
entirely to the closed nature of the circuit. This mechanism is
responsible for the function $f_{0}(\varepsilon, \mathbf{r})$, which
differs substantially from the Fermi function. Therefore, one of the
goals of the present paper is to identify a case in which such a problem
is encountered. Such a problem is encountered in a submicron-thick
thermoelement which is connected to an external circuit.

It is shown in Ref.~\onlinecite{GL92PSS} and~\onlinecite{GLY92} that in 
a submicron film or layer we can ignore the bulk process of energy
relaxation. The effectiveness of the electron-electron collisions is
then governed by the ratio of two frequencies: the frequency of the
electron-electron collisions and the surface energy relaxation
frequency. Since in the case of submicron layers the latter exceeds
greatly the energy relaxation frequency,\cite{GL90} the  
electron-electron collisions are generally ineffective.\cite{foot1} 
Therefore, ``mixing'' of the electron fluxes with a fixed energy 
(partial current) disappears in the bulk and the macroscopic
characteristics should then depend strongly on the actual shape of the
symmetric part of the distribution function. This means that in such a
closed thermoelectric circuit the symmetric part of the electron
distribution function cannot, for fundamental reasons, be approximated
by a Maxwellian function with an effective temperature (which is
frequently done in the case of bulk samples\cite{L67}).

We shall consider a conducting semiconductor film of submicron thickness
$2a$, whose one surface at $x = -a$ is in contact with a heater held at
a temperature $T_{1}$ and the other surface at $x = +a$ is in contact
with a refrigerator held at $T_{2}$. A thermoelectric current can flow
along the $x$ axis when the contacts are closed. For simplicity, we
shall consider only the case in which $T_{1} - T_{2} \ll T_{1}, T_{2}$ 
and then in the first approximation with respect to $\alpha = (T_{1} - 
T_{2})/T$ [where $T = (T_{1} + T_{2})/2]$ we find that the distribution
function is described by the following system of equations:
\begin{equation}
\begin{gathered}
\frac{p}{3m} \frac{\partial f_{1}(\varepsilon, x)}{\partial x} = 0,\\
\frac{1}{\tau(\varepsilon, x)} f_{1}(\varepsilon, x) = - \frac{p}{m}
\left[
\frac{\partial f_{0}(\varepsilon, x)}{\partial x} + 
eE(x)\frac{f_{0}(\varepsilon, x)}{\partial\varepsilon} 
\right] .
\end{gathered}
\label{eq:glt93:f0f1}
\end{equation}
Here $e$, $m$, and $p$ are the charge, mass, and electron momentum;
$\tau(\varepsilon, x)$ is the momentum relaxation time; $E(x)$ is the
electric field intensity; and $\varepsilon$ and $x$ are the energy and 
coordinate of the current carriers.

In the derivation of the system~(\ref{eq:glt93:f0f1}) it is assumed that 
the electron gas is nondegenerate, that the collisions of electrons with
the scattering centers are quasielastic, and that the volume collision
integrals, which represent the energy relaxation in a submicron film,
are unimportant. We can then write the electron distribution
function $f(p, \mathbf{r}, t)$ in the form
\begin{equation}
f(p, \mathbf{r}, t) = f_{0}(\varepsilon, \mathbf{r}) +
f_{1}(\varepsilon, \mathbf{r}) \frac{\mathbf{p}}{p} ,
\label{eq:glt93:fprt}
\end{equation}
where $f_{0}(\varepsilon, \mathbf{r})$ and 
$f_{1}(\varepsilon, \mathbf{r}) (\mathbf{p}/p)$ are, 
respectively, the symmetric part and the anisotropic part of the
distribution function such that $|f_{1}| \ll f_{0}$.

We can easily see that the field $E(x)$ which is contained in 
Eq.~(\ref{eq:glt93:f0f1}) is independent of $x$ if $\alpha \ll 1$. In 
the approximation linear in $\alpha$, the system~(\ref{eq:glt93:f0f1}) 
then reduces to 
\begin{equation}
\frac{\partial^{2}f_{0}(\varepsilon, x)}{\partial x^{2}} = 0.
\label{eq:glt93:d2}
\end{equation}
The distribution function $f_{0}(\varepsilon, x)$ in this case can 
easily be sought in the form 
\begin{equation}
f_{0}(\varepsilon, x) = 
\exp
\left(
\frac{\mu - \varepsilon}{T}
\right) 
\,
\left[
1 + \Psi(\varepsilon, x) \alpha
\right] ,
\label{eq:glt93:f0}
\end{equation}
where $\mu$ is the chemical potential of electrons at a temperature $T$.
The normalization condition then gives
\begin{equation}
\int_{0}^{\infty} d\varepsilon\, g(\varepsilon)  \Psi(\varepsilon, x)
\exp\left(-\frac{\varepsilon}{T}\right)
\label{eq:glt93:int}
\end{equation}
Substituting Eq.~(\ref{eq:glt93:f0}) into Eq.~(\ref{eq:glt93:d2}), we
find the following expression for $\Psi(\varepsilon, x)$:
\begin{equation}
\Psi(\varepsilon, x) = C_{1}(\varepsilon, x)\left[x + 
C_{2}(\varepsilon)\right] .
\label{eq:glt93:Psi}
\end{equation}

The functions $C_{1}(\varepsilon)$ and $C_{2}(\varepsilon)$ represent 
the boundary conditions at the $x = \pm a$ surfaces.

We shall now formulate these boundary conditions. As usual (see 
Ref.~\onlinecite{BBG84}), we assume that between the semiconductor
and the constant-temperature chamber (for example, in the $x = a$ plane)
there is a transition layer of thickness $2\delta$, under the assumption
that the function $f_{0}(\varepsilon, x)$ is continuous at the points
$x = a + \delta$.

The existence of surface energy relaxation mechanisms mentioned above
corresponds to the case in which the first equation in 
system~(\ref{eq:glt93:f0f1}), which describes the distribution function
in a transition layer, contains collision integrals representing energy
transfer from electrons in a given particle flux to the
constant-temperature system whose temperature is $T_{2}$. The absence of
such collision integrals in the case of the submicron film and their
presence in the transition layer should not be regarded as surprising,
since only a peculiar behavior of these integrals (in the limit $\delta 
\to 0$) can give rise to a nonzero surface energy relaxation rate.
Clearly, inclusion of these collision integrals corresponds to 
``mixing'' of the partial fluxes, which facilitates the formation of a
Maxwellian distribution $f_{0}(\varepsilon, x)$ in the semiconductor
near the $x = a - \delta$ surface. In addition to this interaction of
electrons with the constant-temperature system during the flow of a
current, there is another mechanism for the energy exchange between the
electron gas in the film and the constant-temperature chamber (which is
effected by the current itself. Since we are interested in the case
where $f_{0}(\varepsilon, x)$ resembles the Maxwellian shape as little
as possible, we assume that in the case of the transition layer (and the
submicron film) there are no collision integrals\cite{foot2} in the 
first equation in system~(\ref{eq:glt93:f0f1}). It is important to
stress that the mechanism for the energy transfer, which is associated
with the flow of a current, operates even when the total current $j$ is
zero. In fact, it does not follow from $j = 0$ that the partial current
$j(\varepsilon, x)$ vanishes. The partial currents therefore effect the
energy exchange between the electron gas in the film and the
constant-temperature chamber, even when the circuit is open.

Under the assumptions made above it follows from 
system~(\ref{eq:glt93:f0f1}) that
\begin{equation}
j(\varepsilon, x = a - \delta) = j(\varepsilon, x = a + \delta)
\label{eq:glt93:cont}
\end{equation}
where the partial current $j(\varepsilon, x)$ is given by the
expressions\cite{BBG84}
\begin{equation}
j(\varepsilon, x) = - \frac{2e}{3m} \varepsilon g(\varepsilon) 
\tau(\varepsilon) 
\left(
\frac{\partial f_{0}}{\partial x} + eE 
\frac{\partial f_{0}}{\partial\varepsilon} 
\right) .
\label{eq:glt93:jex}
\end{equation}
Here $g(\varepsilon)$ is the density of electron states. The total 
current density $j$ is defined in terms of $j(\varepsilon, x)$ as 
follows:
\begin{equation}
j = \int_{0}^{\infty}d\varepsilon\, j(\varepsilon, x) .
\end{equation}

Substituting in $j(\varepsilon, x = a + \delta)$ the parameters of the
transition layer, replacing the derivative $\partial f_{0s}/\partial x$ 
($f_{0s}$ is the symmetric part of the distribution function in the
transition layer) by $[f_{0}(\varepsilon, x = a + \delta) - 
f_{0}(\varepsilon, x = a - \delta)]/\delta$, and assuming that $\delta$ 
approaches zero, we find from Eq.~(\ref{eq:glt93:jex}) the following
expression (such a procedure is described in greater detail in 
Ref.~\onlinecite{BBG84}):
\begin{multline}
j(\varepsilon, x = a) =
\frac{\sigma_{s}(\varepsilon)}{\sigma_{s}} j -\xi_{s}(\varepsilon)
\exp\left(\frac{\varepsilon - \mu(T)}{T}\right)
\left[
\exp\left(\frac{\mu(T_{2} - \varepsilon}{T_{2}}\right)
- f_{0}(\varepsilon, x = a)
\right] +\\
\frac{\sigma_{s}(\varepsilon)}{\sigma_{s}}
\int_{0}^{\infty}d\varepsilon\, \xi_{s}(\varepsilon) 
\exp\left(\frac{\varepsilon - \mu(T)}{T}\right) 
\left[
\exp\left(\frac{\mu(T_{2} - \varepsilon}{T_{2}}\right) 
- f_{0}(\varepsilon, x = a)
\right] .
\label{eq:glt93:jexa}
\end{multline}
Here
\begin{equation}
\sigma_{s}(\varepsilon) = \lim_{\delta \to 0}
\frac{2e^{2}\varepsilon g_{s}(\varepsilon)\tau_{s}(\varepsilon)
\exp\left(\frac{\mu(T) - \varepsilon}{T}\right)}{3mT} ,
\label{eq:glt93:sigmase}
\end{equation}
\begin{equation}
\xi_{s}(\varepsilon) = \lim_{\delta \to 0}
\frac{2eg_{s}(\varepsilon)\tau_{s}(\varepsilon)
\exp\left(\frac{\mu(T) - \varepsilon}{T}\right)}{3m\delta} ,
\label{eq:glt93:xise}
\end{equation}
and
\begin{equation}
\sigma_{s} = \int_{0}^{\infty}d\varepsilon\, \sigma_{s}(\varepsilon)
\label{eq:glt93:sigmas}
\end{equation}
are the parameters of the transition layer, and $g_{s}(\varepsilon)$ and 
$\tau_{s}(\varepsilon)$ are the density of the electron states and the 
relaxation time in a transition layer.

Substituting in Eq.~(\ref{eq:glt93:jexa}) (and in a relation similar to
it, at $x = -d$) the function $f_{0}(\varepsilon, x)$ in the 
form~(\ref{eq:glt93:f0}) and~(\ref{eq:glt93:Psi}), we find, in a linear 
approximation in $\alpha$
\begin{multline}
\frac{\sigma(\varepsilon)}{\sigma} j_{0} - 
\xi(\varepsilon)C_{1}(\varepsilon) + 
\frac{\sigma(\varepsilon)}{\sigma} \int_{0}^{\infty}d\varepsilon\,
\xi(\varepsilon) C_{1}(\varepsilon)  =
\frac{\sigma_{s}(\varepsilon)}{\sigma_{s}} j_{0}
-\xi_{s}(\varepsilon)
\left[
-C_{1}(\varepsilon) a \pm C_{2}(\varepsilon) + \frac{3}{4} - 
\frac{\sigma_{s}(\varepsilon)}{2T}
\right] +
\frac{\sigma_{s}(\varepsilon)}{\sigma_{s}} 
\int_{0}^{\infty}d\varepsilon\, \xi_{s}(\varepsilon) \times \\
\left[
-C_{1}(\varepsilon) a \pm C_{2} + \frac{3}{4} - \frac{\varepsilon}{2T}
\right] ,
\label{eq:glt93:equality}
\end{multline}
\begin{equation}
\sigma(\varepsilon) = 
\frac{2e^{2}\varepsilon g(\varepsilon) \tau(\varepsilon)
\exp\left[\frac{\mu(T) - \varepsilon}{T}\right]}{3mT}  ,
\label{eq:glt93:sigmae}
\end{equation}
\begin{equation}
\xi(\varepsilon) =
\frac{2e\varepsilon g(\varepsilon) \tau(\varepsilon)
\exp\left[\frac{\mu(T) - \varepsilon}{T}\right]}{3m}  ,
\label{eq:glt93:xie}
\end{equation}
\begin{equation}
\sigma = \int_{0}^{\infty} d\varepsilon\, \sigma(\varepsilon) ,
\qquad
j_{0} = \frac{j}{a} .
\label{eq:glt93:sigma}
\end{equation}

For simplicity we assume that the functions $\sigma_{s}(\varepsilon)$ 
and $\xi_{s}(\varepsilon)$ differ by a constant [as in the case of the 
functions $\sigma(\varepsilon)$ and $\xi(\varepsilon)$]. From 
Eq.~(\ref{eq:glt93:equality}) we can then easily determine 
$C_{1}(\varepsilon)$ and $C_{2}(\varepsilon)$:
\begin{equation}
C_{2}(\varepsilon) = C_{2}^{(0)} = \mathrm{const} ,
\label{eq:glt93:C2e}
\end{equation}
\begin{multline}
C_{1}(\varepsilon) =
\frac{\xi_{s}(\varepsilon)}{\xi(\varepsilon) + a\xi_{s}(\varepsilon)}
\left(
\frac{3}{4} - \frac{\varepsilon}{2T}
\right) +
C_{1}^{(0)} \frac{\sigma(\varepsilon)}{\xi(\varepsilon + 
a\xi_{s}(\varepsilon))} + \\
C_{1}^{(1)} \frac{\sigma_{s}(\varepsilon)}{\xi(\varepsilon + 
a\xi_{s}(\varepsilon))} +
\frac{\sigma_{s}\sigma(\varepsilon) - 
\sigma\sigma_{s}(\varepsilon)}{\sigma\sigma_{s}
\left[
\xi(\varepsilon + a\xi_{s}(\varepsilon))
\right]} j_{0} ,
\label{eq:glt93:C1e}
\end{multline}
where
\begin{equation}
\begin{gathered}
C_{1}^{(0)} = \frac{1}{\sigma}\int_{0}^{\infty} 
d\varepsilon\,\xi(\varepsilon)C_{1}(\varepsilon) ,\\
C_{1}^{(1)} = \frac{1}{\sigma}
\left[
\int_{0}^{\infty} d\varepsilon a\xi_{s}(\varepsilon) C_{1}(\varepsilon)
+ \frac{1}{2T} \int_{0}^{\infty}d\varepsilon\,\xi_{s}(\varepsilon) -
\frac{3}{4}\int_{0}^{\infty}d\varepsilon\,\xi_{s}(\varepsilon)
\right] .
\label{eq:glt93:C10C11}
\end{gathered}
\end{equation}
Since $C_{2}^{(0)}$ does not depend on energy, we find from 
Eq.~(ref{eq:glt93:int}) that $C_{2}^{(0)} = 0$.

The constants $C_{1}^{(0)}$ and $C_{1}^{(1)}$ can be easily found from
system~(\ref{eq:glt93:C10C11}) by substituting $C_{1}(\varepsilon)$.

Since in the temperature approximation we have $f_{0}(\varepsilon, x)
 = \exp\left[\left(\mu\left[T(x)\right] - 
\varepsilon\right)/T(x)\right]$, and since
\begin{equation}
T(x) = T - a \frac{T}{2a} x ,
\label{eq:glt93:Tx}
\end{equation}
it follows that in the case $\alpha\ll 1$ the expression for 
$C_{1}(\varepsilon)$ is 
\begin{equation}
C_{1}(\varepsilon) = \frac{3}{4a} - \frac{\varepsilon}{2aT} ,
\label{eq:glt93:C1ea}
\end{equation}

We can now see that the first term in Eq.~(\ref{eq:glt93:C1ea}) 
represents the influence of the heater and cooler on the distribution
function when their coupling to the electron gas in the submicron
film is due to some of the partial currents induced by the builtin
thermoelectric field and by thermal diffusion. The second and third
terms are due to the drift components of the partial thermoelectric
current in the semiconductor and in the transition layer (compare with
Ref.~\onlinecite{A81}). Finally, the last term is due to the specific
mechanism for the establishment of the distribution function which we
shall now consider.

For simplicity we shall consider a contact between two $n$-type
semiconductors with zero contact potential, with the same electron
densities, but with different energy dependences of the relaxation
times.\cite{foot3} If we ignore the electron-electron collisions and the
collisions with the scattering centers accompanied by energy transfer,
we find that the partial current in the presence of an external voltage,
which induces a current of density $j$, is described by
\begin{equation}
\operatorname{div} j(\varepsilon, x) = 0 .
\label{eq:glt93:div}
\end{equation}

Using Eq.~(\ref{eq:glt93:jex}) and~(\ref{eq:glt93:sigmae}), we obtain 
the following expression from Eq.~(\ref{eq:glt93:div}):
\begin{equation}
\sigma_{1}(\varepsilon)E_{1} = \sigma_{2}(\varepsilon)E_{2}
\label{eq:glt93:E1E2}
\end{equation}
where the index $1$ refers the left-hand semiconductor and the
index $2$ to the right-hand semiconductor. The electron fields $E_{1}$
and $E_{2}$, which are contained in Eq.~(\ref{eq:glt93:E1E2}), can 
easily be expressed in terms of the total density of the current
flowing through the contact:
\begin{equation}
E_{1,2} = j / \sigma_{1,2} .
\label{eq:glt93:E12}
\end{equation}

Using Eq.~(\ref{eq:glt93:E12}), we can write Eq.~(\ref{eq:glt93:E1E2}) 
in the form
\begin{equation}
\frac{\sigma_{1}(\varepsilon)}{\sigma_{1}} - 
\frac{\sigma_{2}(\varepsilon)}{\sigma_{2}} = 0 .
\label{eq:glt93:sigma1sigma2}
\end{equation}

The last equality for different dependences of $\tau_{1}$ and $\tau_{2}$ 
on $\varepsilon$ can be valid only if the distribution functions of the
two semiconductors are non-Maxwellian.

A similar situation occurs also in the absence of a contact if $\tau$ 
depends not only on the carrier energy, but also on the coordinates
[naturally, if $\tau(\varepsilon, x) \ne 
F_{1}(\varepsilon)F_{2}(\varepsilon)$
and, consequently, the coordinate dependence of the relaxation time
determines the energy dependence of the distribution function.

It therefore follows that if the relaxation time in the submicron film
depends in different ways on the carrier energy at different points in
the direction of flow of the current, a new mechanism, which is
responsible for a basically non-Maxwellian distribution function, will
appear.

Turning back to the last term in Eq.~(\ref{eq:glt93:C1e}), we can say 
that it describes this specific mechanism, and that the coordinate
dependence of the relaxation times is related to the different
forms of $\tau(\varepsilon)$ in the submicron layer 
[$\sigma(\varepsilon)$] and at the contact 

If the condition $a\xi_{s}(\varepsilon) \ll \xi(\varepsilon)$ is valid 
for all energies $\varepsilon$, it follows from Eq.~(\ref{eq:glt93:C1e}) 
that
\begin{equation}
C_{1}(\varepsilon) = C_{1}^{(0)} \frac{e}{T} +
\frac{\sigma_{s}\sigma(\varepsilon) - 
\sigma\sigma_{s}(\varepsilon)}{\sigma\sigma_{s}\xi(\varepsilon)} j_{0}
\label{eq:glt93: C1eC10}
\end{equation}

We then see from the Kirchhoff law that $j_{0} = 0$, and
from system~(\ref{eq:glt93:C10C11}) we find that $C_{1}(\varepsilon) = 
0$. Therefore, $\Psi(\varepsilon, x) = 0$.

It is thus clear that if $a\xi_{s}(\varepsilon) \ll \xi(\varepsilon)$ 
then the symmetric part of the distribution function of electrons in a
submicron film is Maxwellian with a temperature equal to $T$.
Such a situation can therefore be called adiabatic.

If, on the other hand, at all values of $\varepsilon$ we have 
$a\xi_{s}(\varepsilon) \gg \xi(\varepsilon)$, then
\begin{equation}
C_{1}(\varepsilon) = \frac{3}{4a} - \frac{\varepsilon}{2aT} + 
C_{1}^{(1)} \frac{\sigma_{s}(\varepsilon)}{a\xi_{s}(\varepsilon)} +
\frac{\sigma_{s}\sigma(\varepsilon - 
\sigma\sigma_{s}(\varepsilon))}{a\sigma\sigma_{s}\xi_{s}(\varepsilon)} 
j_{0} .
\label{eq:glt93:C1eb}
\end{equation}

Since the case $a\xi_{s}(\varepsilon) \ll \xi(\varepsilon)$ has been 
reduced to the adiabatic case (thermal insulation of a film from the
beater and cooler), the case $a\xi_{s}(\varepsilon) \ll 
\xi(\varepsilon)$ can be naturally called isothermal (representing the
ideal thermal coupling between electrons in the submicron film with the
heater and cooler). However, it then follows from 
Eq.~(\ref{eq:glt93:C1eb}) that the symmetric part of the distribution
function of electrons is very far from Maxwellian. In the isothermal
case the distribution function becomes Maxwellian only if the load
resistance approaches infinity and the current $j$ approaches
correspondingly zero. In fact, if $j = 0$, it follows from 
Eq.~(\ref{eq:glt93:C1eb}), with allowance for 
Eq.~(\ref{eq:glt93:C10C11}), that $C_{1}(\varepsilon) = (3/4a) - 
(\varepsilon/2al)$, which - as pointed out above - corresponds to a
Maxwellian function which depends on the coordinates in accordance with
the law (\ref{eq:glt93:Tx}).

The maximum deviation of $f_{0}(\varepsilon, x)$ from the Maxwellian
function occurs, as demonstrated by Eq.~(\ref{eq:glt93:C1e}), when 
$a\xi_{s}(\varepsilon) \approx \xi(\varepsilon)$ and then the energy 
dependences represented by the two functions should be very different.
It should be pointed out that since $a$ is an independent parameter, a
change in the thickness of the semiconductor film may alter the
relationships between $a\xi_{s}(\varepsilon)$ and $\xi(\varepsilon)$, 
which in turn may modify the symmetric part of the distribution 
function.

We have thus shown that the symmetric part of the distribution function
of electrons in a submicron film placed between a heater and a cooler
may be far from Maxwellian. Therefore, in calculations of the
thermoelectric emf and of the thermoelectric current it is necessary to
formulate a new approach which does not rely on such concepts as
temperature and chemical potential. Such an approach will be developed
in a separate paper.

The authors are deeply grateful to V.~I.~Perel', for valuable comments.

Translated by A. Tybulewicz


\begin{references}
\bibitem{SL69}
J.~Shah and~R.~C.~C.~Leite,
Phys.~Rev.~Lett. {\bfseries 22}, 1304 (1969).
\bibitem{GS72}
Yu.~G.~Gurevich and S.~I.~Shevchenko, 
Zh. Eksp. Teor. Fiz. {\bfseries 62}, 806 (1972) 
[Sov. Phys. JETP {\bfseries 35}, 426 (1972)].
\bibitem{VG78}
A.~I.~Vakser and Yu.~G.~Gurevich, 
Fiz. Tekh. Poluprovodn. {\bfseries 12}, 82 (1978) 
[Sov. Phys. Semicond. {\bfseries 12}, 46 (1978)].
\bibitem{BZK82}
R.~Baltrameyunas, A.~Zhukauskas, and E. Kuokshtis, 
Zh. Eksp. Teor. Fiz. {\bfseries 83}, 1215 (1982) 
[Sov. Phys. JETP {\bfseries 56}, 693 (1982)].
\bibitem{GY87}
Yu.~G.~Gurevich and V.~B.~Yurchenko, 
Bulg. J. Phys. {\bfseries 14}, 52 (1987).
\bibitem{GL92PR}
Yu.~G.~Gurevich and G.~N.~Logvinov, 
Phys. Rev. B {\bfseries 46}, 15516 (1992).
\bibitem{BBG84}
F.~G.~Bass, V.~S.~Bochkov, and Yu.~G.~Gurevich, 
{\itshape Electrons and Phonons in Bounded Semiconductors\/} 
[in Russian], Moscow (1984).
\bibitem{IMSK87}
I.~F.~Itskovich, M.~V.~Moskalets, R.~I.~Shekhter, and I.~O.~Kuiik, 
Fiz. Nizk. Temp. {\bfseries 13}, 1034 (1987) 
[Sov. J. Low Temp. Phys. {\bfseries 13}, 588 (1987)].
\bibitem{GY89}
Yu.~G.~Gurevich and V.~B.~Yurchenko, 
Solid State Cottimuni. {\bfseries 72}, 1057 (1989).
\bibitem{GL92PSS}
Yu.~G.~Gurevich and G.~N.~Logvinov, 
Phys. Status Solidi B {\bfseries 170}, 247 (1992).
\bibitem{BG75}
F.~G.~Bass and Yu.~G.~Gurevich, 
{\itshape Hot Electrons and Strong Electromagnetic Waves its
Semiconductor and Gas-Discharge Plasmas\/} (in Russian],
Moscow (1975).
\bibitem{L65}
I.~B.~Levinson, 
Fiz. Tverd. Tela (Leningrad) {\bfseries 7}, 2879 (1965) 
[Sov. Phys. Solid State {\bfseries 7}, 2336 (1966)].
\bibitem{GLY92}
Yu.~G.~Gurevich, G.~N.~Logvinov, and V. B. Yurchenko, 
Fiz. Tverd. Tela (St. Petersburg) {\bfseries 34}, 1666 (1992) 
[Sov. Phys. Solid State {\bfseries 34}, 886 (1992)].
\bibitem{A81}
A.~I.~Ansel'm, 
{\itshape Introduction to Semiconductor Theory,\/} 
Mir, Moscow and Prentice-Hall, Englewood Cliffs, NJ (1981).
\bibitem{GY91}
Yu.~G.~Gurevich and V.~B.~Yurchenko, 
Fiz. Tekh. Poluprovodn. {\bfseries 25}, 2109 (1991) 
[Sov. Phys. Semicond. {\bfseries 25}, 1268 (1991)].
\bibitem{GL90}
Yu.~G.~Gurevich and G.~N.~Logvinov, 
Fiz. Tekh. Poluprovodn. {\bfseries 24}, 1715 (1990) 
[Sov. Phys. Semicond. {\bfseries 24}, 1071 (1990)].
\bibitem{foot1}
At very high electron densities, when the electron -electron collisions are
important even in submicron films, we can use a theory of thermoelectricity
developed in Ref.~\onlinecite{GL92}.
\bibitem{GL92}
Yu.~G.~Gurevich and G.~N.~Logvinov, 
Fiz. Tekh. Poluprovodn. {\bfseries 26}, 1945 (1992) 
[Sov. Phys. Semicond. {\bfseries 26}, 1091 (1992)].
\bibitem{L67}
I.~B.~Levinson, 
{\itshape Author's Abstracts of Doctoral Thesis\/} [in Russian], 
Leningrad (1967).
\bibitem{foot2}
The case considered here corresponds to 
$\xi = \eta_{\mathrm{ext}} = \eta_{\mathrm{int}} = 0$ in 
Ref.~\onlinecite{BBG84}. One of the possible realization of our
calculations is discussed in Ref.~\onlinecite{RGK76}. No problems are
encountered in allowing for the finite nature of the quantities 
$\xi$, $\eta_{mathrm{ext}}$, or $\eta_{\mathrm{int}}$.
\bibitem{RGK76}
E.~I.~Rashba, Z.~S.~Gribnikov, and V.~Ya.~Kravchenko, 
Usp. Fiz. Nauk {\bfseries 119}, 3 (1976) 
[Sov. Phys. Usp. {\bfseries 19}, 361 (1976)]. 
\bibitem{foot3}
Such a case may be encountered, for example, if a large number of
neutral impurities is added to one of the semiconductors.
\end{references}
\end{document}